\documentclass[two column,rsi,amsmath,amssymb,superscriptaddress,pre] {revtex4-1}
\setcitestyle{super}

\usepackage{graphicx}
\usepackage[sort&compress]{natbib}
\usepackage{hhline}
\usepackage{soul}
\usepackage{xcolor}
\usepackage{array}
\usepackage{bm}

\usepackage[normalem]{ulem}
\usepackage{hyperref}

\usepackage[thinlines]{easytable}

\begin{document}
\setcounter{totalnumber}{3}
\renewcommand{\thetable}{\arabic{table}}
\newcolumntype{P}[1]{>{\centering\arraybackslash}p{#1}}

\title{Modeling Interfacial Electron Transfer using Path Integral Molecular Dynamics}

\author{Yoonjae Park}
 \affiliation{Department of Chemistry, Massachusetts Institute of Technology, Cambridge, Massachusetts 02139, USA}

\author{Adam P. Willard}
 \email{awillard@mit.edu}
 \affiliation{Department of Chemistry, Massachusetts Institute of Technology, Cambridge, Massachusetts 02139, USA}

\date{\today}
\vspace{0mm}

\begin{abstract}
Outer sphere electron transfer rates can be calculated from simulation data by sampling the equilibrium statistics of the canonical reaction coordinate -- the vertical energy gap. 
For these calculations, electron transfer is typically represented by an instantaneous change in the atomic partial charges. 
In this manuscript, we present an implementation of this procedure that utilizes an explicit path-integral representation of the transferring electron. 
We demonstrate our methodology by combining path integral molecular dynamics and Marcus-Hush-Chidsey theory to calculate the rate of electron transfer from a Ferrocyanide complex to a gold electrode. 
We consider the dependence of this rate on electron transfer distance and applied potential. 
We find that when the electron is represented explicitly via path integral molecular dynamics, as opposed to implicitly via fixed atomic partial charges, the rates and thermodynamics are more consistent with experimental findings. 
We then apply our methodology to explore the role of bridging spectator cations in modifying electron transfer rates.
We find, once again, that the path integral approach produces specific cation effects that are more consistent with experiment than those in which the transfering electron is represented implicitly.
\end{abstract}

\maketitle

\section{Introduction}

In molecular dynamics simulations of outer sphere electron transfer, the transferring electron is often represented implicitly, in terms of its effect on the nuclear partial charges. This approach omits the effects of electronic fluctuations on solvent reorganization energies and electron transfer rates. 
In this manuscript, we use classical molecular dynamics simulation with a path integral-based representation of a transferring electron to study interfacial electron transfer. 
By comparing to the standard approach, we find that an explicit representation of the electron yields more accurate predictions for electron transfer rates.
We extend our approach to investigate the effects of spectator cations on electron transfer rates. 
Our results indicate that the observed spectator cation effect, i.e., an increase in electron transfer rate with increasing cation size, is due to the ion’s effect on the relative stability of the reduced and oxidized state, and not (as often speculated) by its influence on the solvent reorganization energy.

Most electrochemical technologies require the transfer of electrons across the electrode-solution interface.
The molecular mechanisms that underlie these electron transfer processes are very often poorly understood, which is hindering our ability to optimize device performance and efficiency. 
There are two broad categories of interfacial electron transfer processes: the highly coupled inner-sphere electron transfer and the weakly coupled outer-sphere electron transfer.  
Here, we limit our focus to the latter, in which an electron tunnels between a solvated redox species to an electrode.
This type of outer-sphere electron transfer is believed to dominate current flow in a variety of electrochemical systems, particularly under conditions where direct adsorption is limited. \cite{sinha2022solventmediated-8c8, zhang2023anomalous-014, liu2021adiabatic-d1b, qin2023cationinduced-b6c} 

Thermal fluctuations of the interfacial electrostatic environment are known to play a crucial role in facilitating the process of outer-sphere interfacial electron transfer.
However, the specific roles that solvent molecules and electrolyte species play in mediating electron transfer events remain experimentally inaccessible. 
Therefore, our current molecular-level understanding of these processes is primarily derived from a combination of theory, electronic structure calculation, and atomistic simulation. 

Marcus theory provides a general framework for computing the rates of outer-sphere electron transfer processes in condensed-phase systems.
Within this framework, thermodynamic parameters, such as the solvent reorganization energy, reaction free energy, and the activation energy, can be calculated from the results of equilibrium molecular dynamics simulations of the reactant and product states.
More specifically, the statistics of the vertical energy gap, $\Delta E$ (\textit{i.e.}, the difference in potential energy between the reactant and product states at fixed nuclear position) are compiled to construct the diabatic free energy surfaces – the so-called Marcus parabolas, as illustrated in Fig.~\ref{system}. 
This general approach to studying electron transfer processes therefore requires (\textbf{1}) an accurate model of the reactant and product states, and (\textbf{2}) a robust approach for defining $\Delta E$.

The standard method for meeting these requirements is referred to as the identity exchange (IE) scheme. 
In this scheme, the reactant and product states are represented by classical point-charge force fields that differ only in the distribution of atomic partial charges. 
As such, the transferring electron is described implicitly via the specific arrangement of atomic charges.
In the earliest implementations of the IE scheme, the entire electron transitions between two ionic centers, \cite{kim2021interfacial-b8e, willard2008water-4a4} however, more recent implementations have distributed the transferring electron charge across an entire molecular complex. \cite{huang2021cationdependent-d65, tiwari2016reactive-af4} 
While this approach is both straightforward to implement and computationally efficient, it neglects to account for the spatial fluctuations of that transferring electron. The thermodynamic consequences of these fluctuations have not yet been broadly characterized. 

Here, we introduce a method for modeling outer-sphere electron transfer in which the electron is described explicitly, as a classical ring-polymer.
In this method, the molecular system (e.g., everything except for the transferring electron) is modeled with the same basic point-charge force fields that are used in the IE scheme.
The ring-polymer electron and the molecular system are co-evolved using standard path integral molecular dynamics (PIMD). 
Our PIMD scheme allows us to account for the effects of electronic fluctuations in the reorganization energy, reaction free energy, and electron transfer activation energy. 
While similar PIMD methods have been utilized to evaluate the role of electronic fluctuations in exciton dynamics \cite{park2022, parkjcp2022, park2023prm}, polaron physics \cite{bischak2018tunable-cc3, limmer2020photoinduced-50a}, an electron trapping \cite{remsing2020effective-087} in semiconducting materials,  they have not yet been applied to study interfacial electron transfer. 

The remainder of the manuscript is organized as follows. Section \ref{method} introduces the methodology that we adopt and computational details with the specific system of interest. In Sec.\ref{results}, we report and discuss the key properties of electron transfer computed using both PIMD and IE scheme, and apply the path integral framework to investigate the influence of bridging cation. Finally, Sec.\ref{conclusion} provides a summary and concluding remarks. 

\section{Theoretical Framework}
\label{method}

\subsection{A model of interfacial electron transfer}

We consider a molecular system consisting of a redox species in an electrolyte solution confined between a pair of solid constant potential electrodes, such as depicted in Fig.\ref{system}.
The molecular system is accompanied by a single \textit{active} electron -- modeled as a classical ring polymer -- that is capable of transferring between the redox species and the electrodes. 
Together, the molecular system and the active electron are charge neutral.
The dynamics of the molecular system and the electron ring polymer are simulated with classical molecular dynamics, and the resulting trajectories are analyzed in the context of Marcus theory \cite{blumberger2006, limaye2020understanding-461}.

Model energetics are described by a three-term Hamiltonian, $\mathcal{H}_\mathrm{tot} = \mathcal{H}_{\mathrm{el}} + \mathcal{H}_{\mathrm{mol}} + \mathcal{H}_{\mathrm{int}}$, describing the ring polymer electron, the molecular system, and their interactions, respectively.
The electronic properties are determined by $\mathcal{H}_\mathrm{el}$ and $\mathcal{H}_\mathrm{int}$, which describe the kinetic and potential energy of the electron, respectively.
Specifically,
\begin{equation}
\label{eq:Hel}
\mathcal{H}_{\mathrm{e}} = \frac{\hat{\textbf{p}}_{\mathrm{e}}^2}{2m_\mathrm{e}},
\end{equation}
where $\hat{\textbf{p}}_{\mathrm{el}}$ is the momentum operator of the electron, $m_{\mathrm{e}}$ is the electron mass, and
\begin{equation}
\label{eq:Hint}
\mathcal{H}_{\mathrm{int}} = U_{\mathrm{el-mol}}(\hat{\mathbf{x}}_{\mathrm{el}},\lbrace\mathbf{x}^N\rbrace_{\mathrm{mol}}),
\end{equation}
where $U_\mathrm{el-mol}$ is the interaction potential between the ring-polymer electron and the molecular system, $\hat{\mathbf{x}}_\mathrm{el}$ is the position operator of the electron, and $\lbrace\mathbf{x}^N\rbrace_\mathrm{mol}= \lbrace \textbf{x}_1, \ \textbf{x}_2, \dots , \textbf{x}_N \rbrace$ denotes the positions of the $N$ atoms that comprise the molecular system.
The molecular system properties are determined by the molecular Hamiltonian,
\begin{equation}
\mathcal{H}_\mathrm{mol}=\sum_{i=1}^N \frac{\mathbf{p}_i^2}{2m_i} + U_{\mathrm{mol}}(\lbrace\mathbf{x}^N\rbrace_{\mathrm{mol}}),
\end{equation}
where $\mathbf{p}_i$ and $m_i$ are the momentum and mass of atom $i$, and $U_\mathrm{mol}$ is the interaction potential governing all inter-molecular interactions, formulated as a classical molecular mechanics force field.

\subsection{The electron ring polymer}

To capture the quantum mechanical nature of the electron, we adopt a formalism based on the imaginary time path integral \cite{chandler1981exploiting, ceperley1995path, feynman1, feynman2}. 
In this formalism, the partition function of a molecular system, $\mathcal{Z}_\mathrm{}$, is be written as
\begin{equation}
\begin{aligned}
\mathcal{Z}=\int d^{3N} \lbrace \mathbf{x}^N \rbrace_\mathrm{mol} e^{-\beta U_\mathrm{mol}} \times \mathcal{Z}_\mathrm{el}[\lbrace \mathbf{x}^n \rbrace_\mathrm{el} ],
\end{aligned}
\end{equation}
where the partition function of the active electron, $\mathcal{Z}_\mathrm{el}$, is given by, 
\begin{equation}
\begin{aligned}
\mathcal{Z}_\mathrm{el}[\lbrace \mathbf{x}^n \rbrace_\mathrm{el}]=\int d^{3n} \lbrace \mathbf{x}^n \rbrace_\mathrm{el} e^{-(\mathcal{S}_\mathrm{el} + \mathcal{S}_\mathrm{int})}/\hbar 
\label{partition}
\end{aligned}
\end{equation}
where $\beta^{-1} = k_{\mathrm{B}}T$, and $T$, $k_{\mathrm{B}}$, and $\hbar$ denote the temperature, Boltzmann constant, and reduced Planck's constant, respectively.
The parameter $n$ denotes the number of discretized slices along the imaginary time path. 
%
The path actions for the electron and its interactions are given by
\begin{equation}
\begin{aligned}
\mathcal{S}_{\mathrm{el}} = \int_{\tau = 0}^{\beta \hbar} \mathcal{H}_{\mathrm{el},\tau}, \ 
\mathcal{S}_{\mathrm{int}} = \int_{\tau = 0}^{\beta \hbar} \mathcal{H}_{\mathrm{int},\tau}
\label{Sel}
\end{aligned}
\end{equation}
where the $\tau$-dependent Hamiltonian represents its classical analog at imaginarity time $\tau$. For practical implementation, the path action is discretized into $n$ imaginary time slices, such that the quantum particle is equivalently represented as a classical ring polymer composed of $n$ beads linked by harmonic springs.\cite{habershon2013ring}
The associated path action is given by,
\begin{equation}
\begin{aligned}
\mathcal{S}_{\mathrm{el}} =
\sum_{i=1}^n \frac{m_{\mathrm{e}}n}{2\beta \hbar} (\textbf{x}_{\mathrm{el},i} - \textbf{x}_{\mathrm{el},i+1})^2 
\label{HRP}
\end{aligned}
\end{equation}
where $\textbf{x}_{\mathrm{el},i}$ is the position of $i^{\mathrm{th}}$ bead with $\textbf{x}_{\mathrm{el},n+1} = \textbf{x}_{\mathrm{el},1}$. Similarly, $\mathcal{S}_{\mathrm{int}}$ is determined based on Eq.\ref{eq:Hint}, where the imaginary time path is fully resolved and the resultant action is evaluated as a sum over all time slices. 

The interactions between the electron and classical nuclei is modeled using a pseudo potential in form of a truncated Coulomb potential.
Specifically, the interaction potential between $i^{\mathrm{th}}$ bead of the ring-polymer electron and $j^{\mathrm{th}}$ classical atom in the molecular system is given by, 
\begin{equation}
\begin{aligned}
U_{\mathrm{el-mol}}^{ij} = \frac{q_i q_j }{4 \pi \varepsilon_0 n \sqrt{\alpha_{j} + |\textbf{x}_{\mathrm{el},i} - \textbf{x}_j |^2}}
\label{pseudo}
\end{aligned}
\end{equation}
where $q$ and $\varepsilon_0$ are the charge and vacuum permittivity, respectively. The pseudopotential parameter, $\alpha$, is chosen based on the charge and characteristic size of each atom,\cite{parrinello1984study, schnitker1987electron, kuharski1988molecular} along with the condition that the electron remains localized around the redox species.

\subsection{Computing diabatic free energy surfaces}
\label{sec:marcus}

\begin {figure*}
\centering\includegraphics [width=12.5cm] {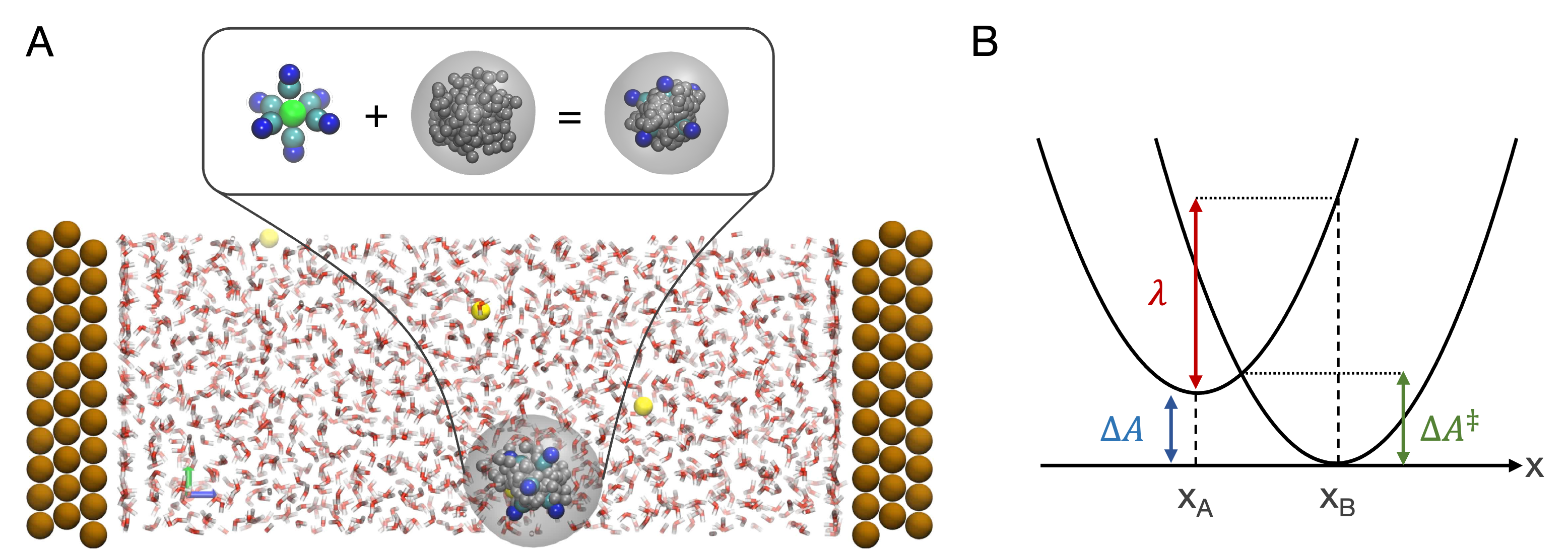}
\caption{
(A) Snapshot of a molecular dynamics simulation illustrating [Fe(CN)$_6$]$^{4-}$ in an aqueous electrochemical cell between two gold electrodes. As shown in the top schematic, an electron$-$represented as a ring polymer (gray)$-$is bound to [Fe(CN)$_6$]$^{3-}$, forming [Fe(CN)$_6$]$^{4-}$. Atoms shown in red, white, brown, green, cyan, and blue represent O, H, Au, Fe, C, and N, respectively, and yellow atoms are K$^+$ ions for system charge neutrality. By convention, the z-axis is defined as perpendicular to the electrode surfaces. The gray shaded areas around the electron are visual guides to emphasize the ring polymer representation of the electron. 
(B) Schematic illustration of diabatic free energy surfaces along a reaction coordinate x labeling key quantities involved in electron transfer -- reorganization energy ($\lambda$), thermodynamic driving force ($\Delta A$), and activation energy ($\Delta A^{\ddagger}$) -- between state A and state B, within the Marcus theory.
In the present study, electron transfer process is considered from state B to state A. 
}
\label{system}
\end{figure*}

Marcus theory establishes that the canonical reaction coordinate for solution-phase outer-sphere electron transfer is the \textit{vertical energy gap}, commonly denoted as $\Delta E$ \cite{blumberger2006, limaye2020understanding-461}.
The vertical energy gap is the instantaneous energy difference between the oxidized and reduced states of the system at fixed nuclear configurations.
The diabatic free energy surfaces are related to the equilibrium statistics of $\Delta E$ through the foundational relationship, $A(\Delta E)=-k_\mathrm{B}T \ln P(\Delta E)$. 
In a system that obeys linear response, this relationship yields diabatic free energy surfaces that are quadratic, otherwise known as \textit{Marcus parabolas}, such as illustrated in Fig.~\ref{system}.

We generate diabatic free energy surfaces by sampling the equilibrium statistics of $\Delta E$ for configurations originating from the oxidized or reduced diabatic states.
For the reduced state diabatic free energy surface, we sample equilibrium configurations where the electron occupies the redox species, computing $\Delta E = E_\mathrm{ox}(\lbrace \mathbf{r}^N \rbrace_\mathrm{mol}) - E_\mathrm{red}(\lbrace \mathbf{r}^N \rbrace_\mathrm{mol})$, where $E_\mathrm{ox}(\lbrace \mathbf{r}^N \rbrace_\mathrm{mol})$ and $E_\mathrm{red}(\lbrace \mathbf{r}^N \rbrace_\mathrm{mol})$ are the potential energies of the system in the oxidized and reduced states, respectively, at fixed nuclear configuration, $\lbrace \mathbf{r}^N \rbrace_\mathrm{mol}$.
More specifically, $E_\mathrm{red}$ is the total system potential energy when the ring polymer electron occupies the redox species, in its original equilibrium configuration, and $E_\mathrm{ox}$ is the total system potential energy with the electron removed from the simulation and the electrode charges adjusted to restore constant potential.
In the latter case, the transferring electron is assumed to equilibrate within the electronic manifold of states within the potentiostatically controlled electrode.

If the equilibrium statistics exhibit Gaussian statistics, then we approximate the full free energy profiles with parabolic fits,
\begin{equation}
\begin{aligned}
A(\Delta E) =-\frac{(\Delta E - \langle \Delta E \rangle )^2}{2 \sigma^2 }
\label{Gaussfit}
\end{aligned} 
\end{equation}
where $\langle \Delta E \rangle$ $\sigma$ are the mean and standard deviation of the vertical energy gap distribution. 
With the $\Delta E$ as the reaction coordinate, the free energy surface of the oxidized state can be constructed analytically from that of the reduced state (or visa versa). 
In the linear response case, the curvature of the oxidized and reduced Marcus parabolas are identical, \textit{i.e.}, $\sigma_{\mathrm{ox}} = \sigma_{\mathrm{red}}$, and their means are related by $\langle \Delta E \rangle_{\mathrm{ox}} = \langle \Delta E \rangle_{\mathrm{red}} - \sigma_{\mathrm{red}}^2 / k_{\mathrm{B}}T$. 
Consequently, the free energy difference between the two states, identified as the thermodynamic driving force, is expressed as $\Delta A = A_\mathrm{ox}({\langle \Delta E \rangle_\mathrm{ox}})- A_{\mathrm{red}}(\langle \Delta E \rangle_{\mathrm{red}})= m_{\mathrm{red}} - \sigma_{\mathrm{red}}^2 / (2 k_{\mathrm{B}}T)$. 
The Marcus parabolas allow us to determine two key kinetic parameters: the reorganization energy, $\lambda = \sigma_{\mathrm{red}}^2 / (2k_{\mathrm{B}}T)$, and the activation energy, $\Delta A^{\ddagger} = (\lambda + \Delta A)^2 / (4\lambda)$.


\subsection{Calculating Heterogeneous Electron Transfer Rate Constant} 
\label{mhctheory}

\begin {figure}
\centering\includegraphics [width=8.6cm] {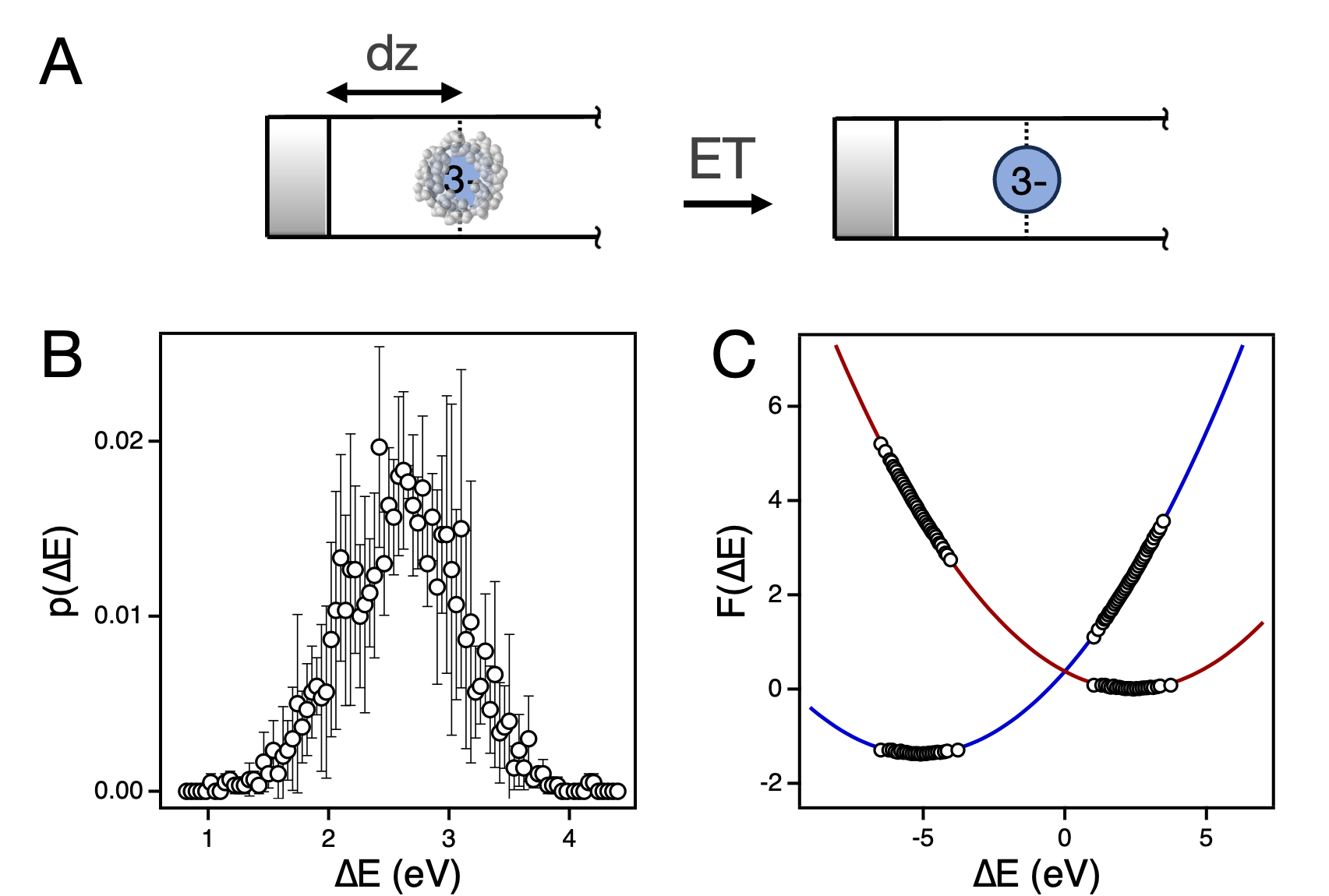}
\caption{(A) Schematic illustrating the inner-sphere electron transfer process via the PIMD scheme. 
In the reduced state (left), the electron occupies the ferricyanide complex, resulting in a total charge 4-, whereas in the oxidized state (right), the electron has merged with the charge distribution of the constant potential electrode 
(B) Probability distribution $p(\Delta E)$ with the values of $\Delta E$ collected from a reduced state. 
(C) The representative free energy surfaces $F(\Delta E)$ where the red and blue curves correspond to the parabolic fits for the reduced and oxidized diabatic free energy surfaces, respectively.
Symbols denote the statistics derived from simulation data.
In panels (B) and (C), $dz=2\,\mathrm{nm}$ where $dz$ is defined as the shortest distance between the center of mass of the anion complex and the first layer of electrode atoms. 
}
\label{etmethod}
\end{figure}

Our computational protocol for computing $\Delta E$ is illustrated in the schematics of Fig.\ref{etmethod}.
In this way, the net charge difference between the reduced and oxidized states is $-e$ and $+e$ for the electrodes and the anion complex, respectively, which we demonstrate in the SI (Fig.S1B).

We base our analysis of interfacial electron transfer kinetics on Marcus–Hush–Chidsey (MHC) theory \cite{kurchin2020marcushushchidsey-b8d, henstridge2012marcushushchidsey-137, zeng2014simple-62d}.
This theoretical framework extends the classical Marcus theory by incorporating the influence of electrode density of states and Fermi-Dirac statistics, making it well suited for modeling interfacial reactions under electrochemical conditions. 
At a given overpotential $\eta$, the rate expression is given by 
%
\begin{equation}
\begin{aligned}
k_{\mathrm{MHC}} 
= \gamma \int_{-\infty}^{\infty} dx \frac{1}{1 + e^{x/k_{\mathrm{B}} T }}
\exp \bigg\{ -\frac{\left( x - \lambda + e \eta \right)^2}{4\lambda k_{\mathrm{B}}T} \bigg\}
\label{kmhc} 
\end{aligned}
\end{equation}
with
\begin{equation}
\begin{aligned}
\gamma = \frac{|K|^2}{\hbar \left( 4\pi \lambda k_{\mathrm{B}} T \right)^{1/2}} 
\label{gamma} 
\end{aligned}
\end{equation}
where $\gamma$ is the pre-exponential factor, incorporating the electronic coupling strength $K$, and the variable of integration, $x$, represents the energy level of electronic states in the electrode relative to the Fermi level. 
The overpotential is given by $\eta = V_\mathrm{ext} - \Delta A$, where $V_\mathrm{ext}$ is the applied electrode potential relative to the potential of zero charge.
In the limit where the reorganization energy $\lambda$ is much larger than the thermal energy $k_{\mathrm{B}}T$, the MHC rate expression can be simplified analytically to yield the following expression,
\begin{equation}
\begin{aligned}
k_{\mathrm{MHC}} 
= \gamma \frac{\sqrt{\pi \lambda ' }}{1 + e^{-\eta '} } \, \mathrm{erfc} \Bigg( \frac{\lambda ' - \sqrt{a + {\eta '}^2}}{2 \sqrt{\lambda '}} \Bigg) 
\label{kmhc2} 
\end{aligned}
\end{equation}
where $a = 1 + \sqrt{\lambda '}$ 
and the primed parameters represent reduced quantities normalized by the thermal energy $k_{\mathrm{B}}T$. 
This expression captures the probabilistic nature of electron occupancy in the electrode and the thermal broadening of energy levels at finite temperature, highlighting that the MHC framework offers a refined theoretical description of electron transfer kinetics at electrode interfaces, particularly under electrochemical conditions.

\subsection{Computational Methodology}
\label{details}

The section above describes a general approach to using PIMD for computing interfacial outer-sphere electron transfer rates.
To demonstrate the utility of this approach, we apply it study a specific system: the outer-sphere electron transfer from a ferrocyanide complex [Fe(CN)$_6$]$^{4-}$ to a gold electrode at the aqueous electrode interface.
This well-studied ET reaction serves as a representative system to validate our method and to highlight the effects of electronic fluctuations on the kinetics and thermodynamics of electron transfer.

\subsection{Atomistic Simulation Details} 
\label{simdetails}

We perform atomistic molecular dynamics simulations to study heterogeneous electron transfer of ferrocyanide in aqueous electrochemical cell, as illustrated in Fig.~\ref{system}. 
The system contains 1658 water molecules, a single [Fe(CN)$_6$]$^{4-}$ anion, and 4 K$^+$ as counterions.
The solution is confined between two parallel electrodes aligned in the $xy$-plane and separated by a distance of 8nm. 
The $z$-axis therefore defines the direction perpendicular to the electrode surfaces. 
Each electrode consists of three layers of atoms arranged in an ideal FCC lattice, with lattice constant $d_\mathrm{Au}=4.17$ (consistent with metallic Au), and the 111 facet exposed to the solution.
The reduced and oxidized states of the redox species are distinguished based on the location of the ring polymer electron.
When the redox species and the electron are spatially separate, the species is considered to be in the oxidized state, \textit{i.e.}, ferricyanide, with a net charge of $-3e$.
When the redox species and the electron are colocalized (bound), the species is in the reduced state, \textit{i.e.}, ferrocyanide, with a net charge of $-4e$.
The snapshot in Fig.~\ref{system} depicts a ferricyanide configuration.
While conceptually simple, this approach reasonably reproduces the solvation structures reported in the literature \cite{huang2021cationdependent-d65} around the anion in both ferri- and ferrocyanide states (Fig.S1A). The electron is quantized by 1000 time slices which is large enough to properly describe the behavior of electron in water where fictitious masses for the beads are set to 1 amu for computational simplicity. 

The dimension of the system is $2.7 \mathrm{nm} \times 2.7 \mathrm{nm} \times 8.0 \mathrm{nm}$, ensuring that the system captures both bulk and interfacial regions \cite{olivieri2021confined-dc9}, and periodic boundary conditions are applied in the $x$ and $y$ directions. For interaction potentials,  SPC/E model \cite{berendsen1987missing-81b} is used for water and metal electrodes are modeled using parameters developed by Heinz \textit{et al.} \cite{heinz2008accurate-8ec}. The parameters for the ferrocyanide and counterions are adopted from the studies in the literature \cite{huang2021cationdependent-d65, prampolini2014structure-8cf}, which has been shown to accurately reproduce the structure of the anion in the aqueous environment. A summary of parameters including pseudopotentials used in the simulations is provided in Supporting Information (Table S1 and S2). 

Simulations were performed in the canonical (NVT) ensemble with Langevin thermostat to control the temperature and a timestep of 1 fs. We employ constant potential method \cite{siepmann1995influence-de2} using the LAMMPS \cite{lammps} ELECTRODE package \cite{ahrensiwers2022electrode-cc9} where the charges of electrode atoms fluctuate in response to the nearby electrostatic environment, resulting in a constant potential between the two electrodes. The positions of the electrode atoms are fixed during the simulations. In addition, the bond lengths and angles of water molecules were constrained using the SHAKE algorithm \cite{ryckaert1977numerical-f67} and long-range electrostatic interactions were treated with the particle-particle particle-mesh algorithm with a real-space cutoff of 13 $\mathrm{\AA}$. All simulations were carried out using LAMMPS.


\subsection{A Comparative Identity Exchange Scheme} 
\label{iescheme}

To benchmark and validate our PIMD scheme, we compare our results to the standard IE scheme.
In the IE scheme, the oxidized and reduced states differ only in the distribution of atomic point charges (\textit{i.e.}, all other force field parameters are identical) and $\Delta E$ is computed by changing the charges at fixed nuclear configuration.
Unlike our PIMD scheme, which explicitly capture the fluctuating effects of quantum delocalization, the IE scheme treats the transitioning electron as a static object, in terms of fixed atomic partial charges.
The IE approach has been widely used in classical molecular simulations for computing vertical energy gaps and reorganization energies within the Marcus framework. 
To facilitate a consistent comparison, the PIMD- and IE-based simulations share an identical set of force field parameters, except for the partial charges on the ferro-cyanide complexes. 
In the IE scheme, $\Delta E$ is computed by switching the values of the atomic charges of the ferri- or ferro-cyanide complex, at fixed nuclear coordinates.
The force field parameters for the IE scheme are specified in Table S3. 



\section{Results and discussions}
\label{results}

\begin {figure*}
\centering
\includegraphics [width=13.5cm] {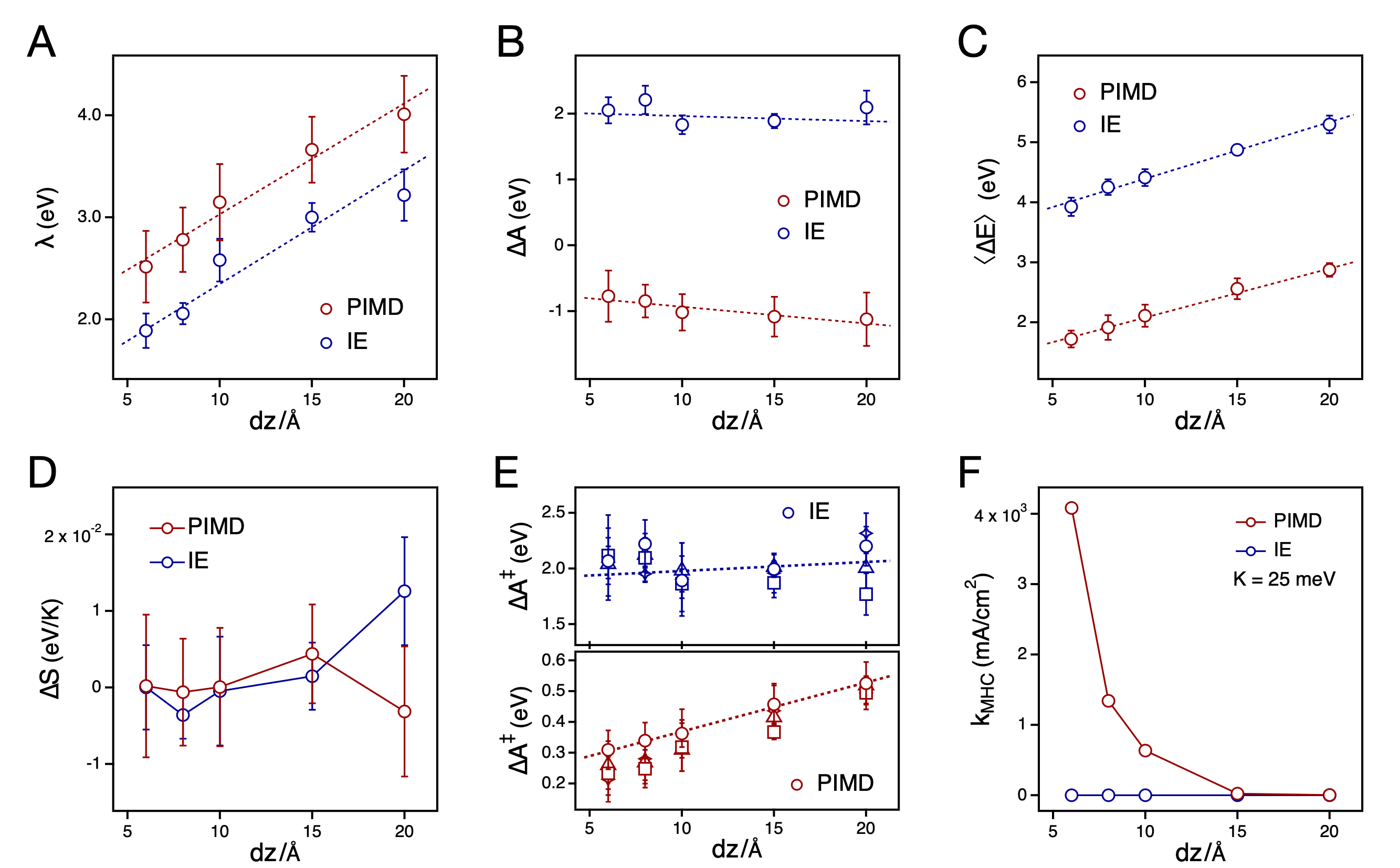}
\caption{The dependence of electron transfer properties on transfer distance, $dz$.
In each panel, we compare the results of PIMD- and IE-based sampling schemes, as plotted in red and blue, respectively.
(A) Reorganization energy, $\lambda$. 
(B) Electron transfer free energy, $\Delta A$.
(C) Mean vertical energy gap $\langle \Delta E \rangle$.
(D) Entropic driving force, $\Delta S_{\mathrm{}}$.
(E) Activation energy $\Delta A^{\ddagger}$. For panel (E), different symbols indicate different temperatures: stars (280K), circles (298K), triangles (320K), and squares (340K).
(F) MHC ET rate constants, $k_{\text{MHC}}$, with constant electronic coupling strength $K = 25\,\text{meV}$ \cite{huang2021cationdependent-d65}. 
Dotted lines indicate linear fits, and solid lines are guides to the eye. 
} 
\label{V0prop} 
\end{figure*}

In this section, we present simulation results focused primarily on calculations of the Marcus-Hush-Chidsey electron transfer rate, $k_\mathrm{MHC}$, along with the termodynamic parameters $\lambda$, $\Delta A$, and $\Delta A^{\ddagger}$, under varying system conditions.
We benchmark our results against experiment, when available, although quantitative agreement is not expected due to the highly idealized nature of our simulation setup (\textit{e.g.}, featureless electrodes, non-polarizable solvent, etc.). 
Despite this, we expect that the general trends we observe are qualitatively reliable.
To evaluate the consequences of modeling the transitioning electron explicitly, we compare the results of our PIMD scheme to those generated with an IE scheme carried out on a practically identical system, as described above in Sec.~\ref{iescheme}.

\subsection{Dependence of $k_\mathrm{MHC}$ on electron transfer distance}


We analyze the statistics of the vertical energy gap, $\Delta E$, to derive key ET properties$-$ including reorganization energy ($\lambda$), activation energy ($\Delta A^\ddagger$), and reaction free energy ($\Delta A$), schematically illustrated in Fig.\ref{system}. 
To investigate the spatial dependence of electron transfer rate on transfer distance, we present these key properties as a function of the separation distance between the redox complex and the electrode surface, denoted as $dz$.
We find that in all of our simulations, the statistics of $\Delta E$ are approximately Gaussian, indicating that the simplifying approximations of Marcus theory (such as discussed in Sec.~\ref{sec:marcus}), can be applied to the analysis of our data.

The width of the Marcus parabola exhibits a systematic trend, with $\sigma_\mathrm{ox}=\sigma_\mathrm{red}$ increasing as the redox species-electrode separation, $dz$, decreases.
This trend leads to a corresponding trend in the reorganization energy, with $\lambda$ decreasing as $dz$ decreases.
As Fig.~\ref{V0prop}A highlights, this trend is observed for both the PIMD and IE schemes, which follow the same trend but differ by an additive constant (arising from differences in the value of $\Delta A$).
This shared distance dependent trend reflects the classical image charge effect, \cite{imagemethod} where closer proximity to the electrode stabilizes charge distributions and compresses the energetics, confirming that our approach captures the expected electrostatic behavior of ET at the interface. 

The dependence of $\Delta A$ on $dz$ is plotted in Fig.~\ref{V0prop}B.
We observe that the two methods yield qualitatively different results.
Although neither approach exhibits significant dependence on $dz$, we find that $\Delta A<0$ in the PIMD simulations (indicating the oxidized state is more stable) 
while $\Delta A>0$ for the IE simulations.
This discrepancy in sign likely arises from the fundamentally different treatment of the transferring electron. In PIMD, the excess electron is explicitly represented as a quantum ring polymer, allowing it to polarize and interact with the environment in a physically consistent way. In contrast, the IE scheme enforces a charge reassignment without accounting for the microscopic solvation or electronic delocalization, potentially overstabilizing the reduced state.
This simplified treatment also contributes to the markedly larger $\Delta E$ values observed in the IE model, as plotted in Fig.~\ref{V0prop}C, reflecting its inability to capture quantum mechanical delocalization of the electron. 

We quantify the entropic contribution to electron transfer, $\Delta S$, by performing a linear fit of the electron transfer free energy $\Delta A$ as a function of temperature, with the slope of yielding $-\Delta S$. 
This quantity reflects thermodynamic contributions from the ensemble of nuclear configurations that contribute to electron transfer.
The dependence of $\Delta S_\mathrm{}$ on $dz$ is plotted in Fig.~\ref{V0prop}D.
We observe that the variation of $\Delta S$ with $dz$ is negligible, and the values obtained from PIMD simulations and the identity exchange scheme are largely consistent with each other. This indicates that entropic effects play only a minor role in the overall driving force, and that the electron transfer process is primarily governed by energetic contributions. 

The effect of ET rate on distance can be intuitively understood by considering the effect of $dz$ on the activation energy, $\Delta A^\ddagger$, as plotted in Fig.~\ref{V0prop}E.
For the IE scheme, the dependence of $\Delta A^{\ddagger}$ on $dz$ is relatively weak, while in contrast, the PIMD results exhibit a more pronounced $dz$ dependence, with $\Delta A^{\ddagger}$ decreasing as the redox center approaches the electrode. 
This trend indicates that at closer distances, enhanced electrostatic interactions and interfacial solvent polarization more effectively stabilize the transition state.
As our data indicates, this trend is observed across a range of different temperatures.

The values of $\lambda$ and $\Delta A$ combine to yield the Marcus-Hush-Chidsey interfacial ET rate constant, $k_\mathrm{MHC}$, as presented in Eq.~\ref{kmhc2}.
The resulting dependence of $k_\mathrm{MHC}$ on $dz$ is plotted in Fig.~\ref{V0prop}F.
The PIMD scheme predicts rate constants with a pronounced distance dependence, increasing significantly as the redox species approaches the electrode.
We observe that the averaged rate constants at separations near $dz \approx 10\mathring{\mathrm{A}}$ are in reasonable agreement with experiment values ($\sim$ 0.1 cm/s). \cite{pajkossy2025diffusionaffected-e78, pajkossy2021dynamic-9b0}
This observation suggests (unsurprisingly) that in the physical system, redox current is largely influenced by species located near the electrode, though not necessarily in direct contact with it. 
We note that since the electronic coupling is assumed to be constant in our analysis, this reported distance dependence is completely due to changes in the thermodynamic parameters $\lambda$ and $\Delta A$.
In contrast, the IE scheme predicts unphysically small rates, effectively vanishing across all distances, with no significant distance dependence.
The contrast between the PIMD- and IE-based rate constants highlights a fundamental distinction between the two approaches.
For the IE-based approach, agreement with experiment evidently relies on the emergence of a strong interfacial electronic coupling, while in the PIMD approach, both coupling and solvent/electronic fluctuations contribute to the large observed electron transfer rates.

When comparing our simulation results with experimental estimates, we find that the computed reorganization energies tend to slightly overestimate the reported values, yet remain within a physically reasonable range. \cite{ghosh2016electrochemical-156, ramírez2021fermi-29b, royea2006comparison-caa} For the activation barrier, the PIMD results fall closer to experimentally plausible values, \cite{kim2016activation-900, huang2021cationdependent-d65} reinforcing the validity of the path-integral framework in capturing realistic interfacial ET energetics. This comparison not only reveals the quantitative differences between the two methods but also highlights scenarios in which the classical IE scheme yields inconsistent or unphysical trends. At the same time, the analysis also identifies properties for which both approaches produce qualitatively similar results, providing insight into when the IE model may offer a reasonable approximation and when a quantum treatment is essential. 
Overall, our results highlight the potential importance of accounting for the effects of electronic fluctuations in calculations of outer-sphere electron transfer rates.

\subsection{Dependence of ET rate on applied electrode potential}

\begin {figure}
\centering\includegraphics [width=8.7cm] {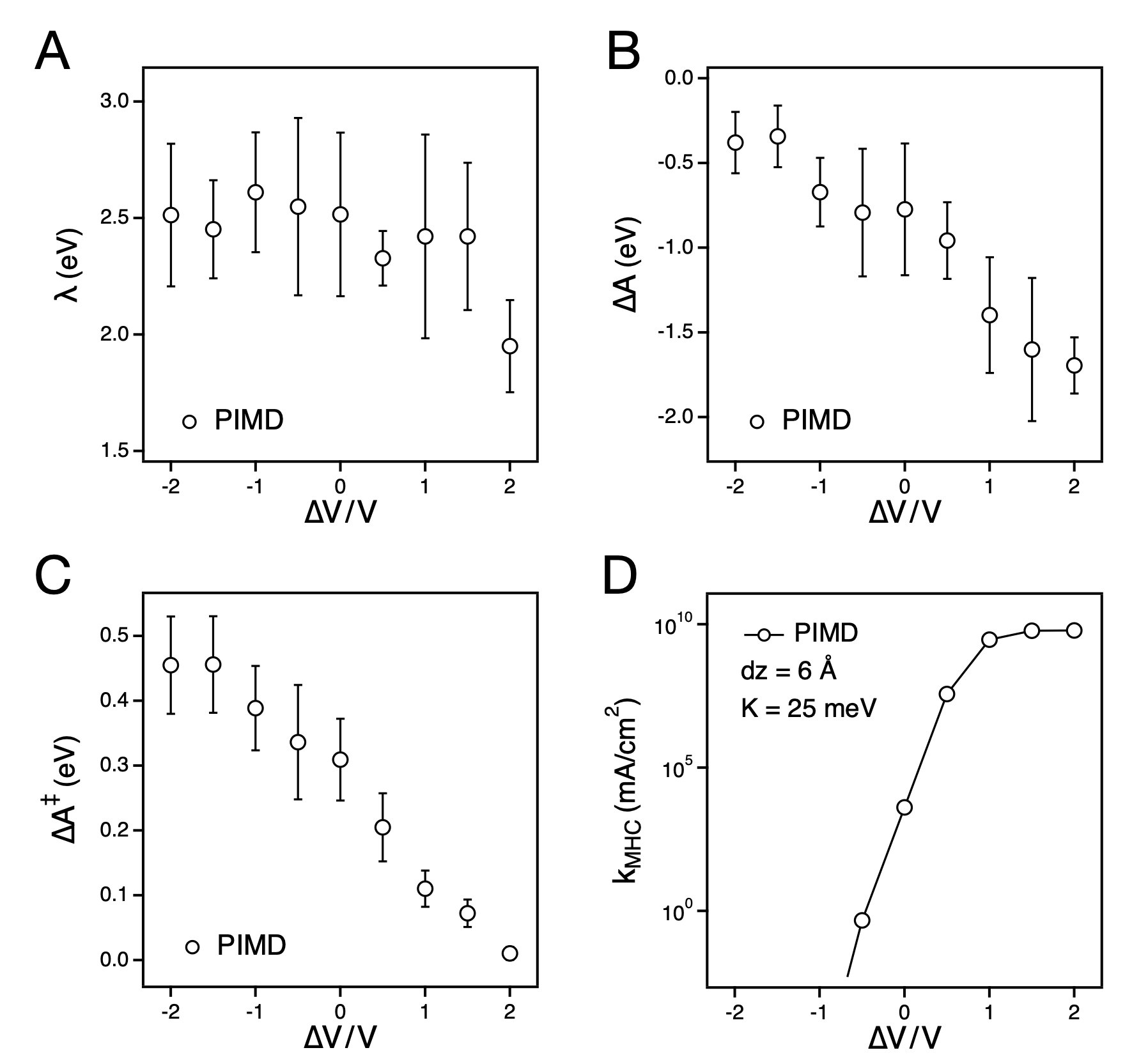} 
\caption{
Response of electron transfer properties to the applied electrode potential $\Delta V$ at a fixed anion–electrode separation of $dz = 6 \mathrm{\AA}$, computed using path integral molecular dynamics. 
(A) Reorganization energy $\lambda$. 
(B) thermodynamic driving force $\Delta A$. 
(C) Activation energy $\Delta A^{\ddagger}$
(D) Resulting ET rate constants $k_{\mathrm{MHC}}$. 
}
\label{z64prop}
\end{figure}

We now examine how the properties that determine $k_\mathrm{MHC}$ are affected by changes in the applied electrode potential as calculated with the PIMD scheme.
To isolate the effects of the electrode potential, we constrain the position of the redox species at $dz=6\mathring{\mathrm{A}}$.
We observe that the reorganization energy shows only a modest dependence on the electrode potential, with a general tendency to decrease as the potential increases, as plotted in Fig.~\ref{z64prop}A. To further examine this behavior, we also carried out calculations using IE scheme, which revealed a similarly weak decreasing trend (Fig.S2). 
This observation is inconsistent with the MHC theory assumption that the reorganization energy is potential independent.
Our results suggest that interfacial electric fields affect solvent structure and dynamics in such a way as to fluctuation amplitudes. 
%

The thermodynamic driving force $\Delta A$ (Fig.\ref{z64prop}B) exhibits a pronounced linear decrease with increasing potential.
This trend reflects the downward shift of the Fermi level relative to the redox species at more positive potentials, which preferentially stabilizes the oxidized state relative to the reduced state. 
Similarly, the activation energy $\Delta A^{\ddagger}$ decreases nearly linearly with increasing $\Delta V$ (Fig.\ref{z64prop}C), consistent with Marcus theory in the regime where $\lambda$ varies modestly and $\Delta A$ dominates the barrier height.

We observe that the overall electron transfer rate, $k_{\mathrm{MHC}}$, computed via Eq.~\ref{kmhc2} and plotted in Fig.~\ref{z64prop}D, exhibits a significant increase with increasingly applied potential, due primarily to the reduction in activation barrier and enhanced thermodynamic driving force.
This behavior aligns with the expected directionality of electron transfer under oxidation bias and further demonstrates that the path integral framework yields physically meaningful rate trends across the electrochemical potential range explored.

\subsection{Influence of Bridging Cations on Interfacial ET}
\label{bridging} 

There is ample evidence that the rate of outer sphere interfacial electron transfer is sensitive to the identity of cation in the supporting electrolyte.
However, the physical origin of this \textit{specific cation effect} remains a topic of debate.
Some hypothesize that these effects arise through the cations' influence on solvent reorganization energy, $\lambda$, \textit{e.g.}, the so-called structure-making and structure-breaking influence on aqueous molecular structure. \cite{ringe2019understanding-24c, zhang2024probing-a4e, huang2021cationdependent-d65} 
Others hypothesize that cations facilitate electron transfer by bridging the gap between the redox species and the electrode, thereby offering a more favorable tunneling environment for the electron. \cite{qin2023cationinduced-b6c} 
Here, we use our PIMD scheme to evaluate these two hypotheses.

\begin {figure}
\centering\includegraphics [width=8.7cm] {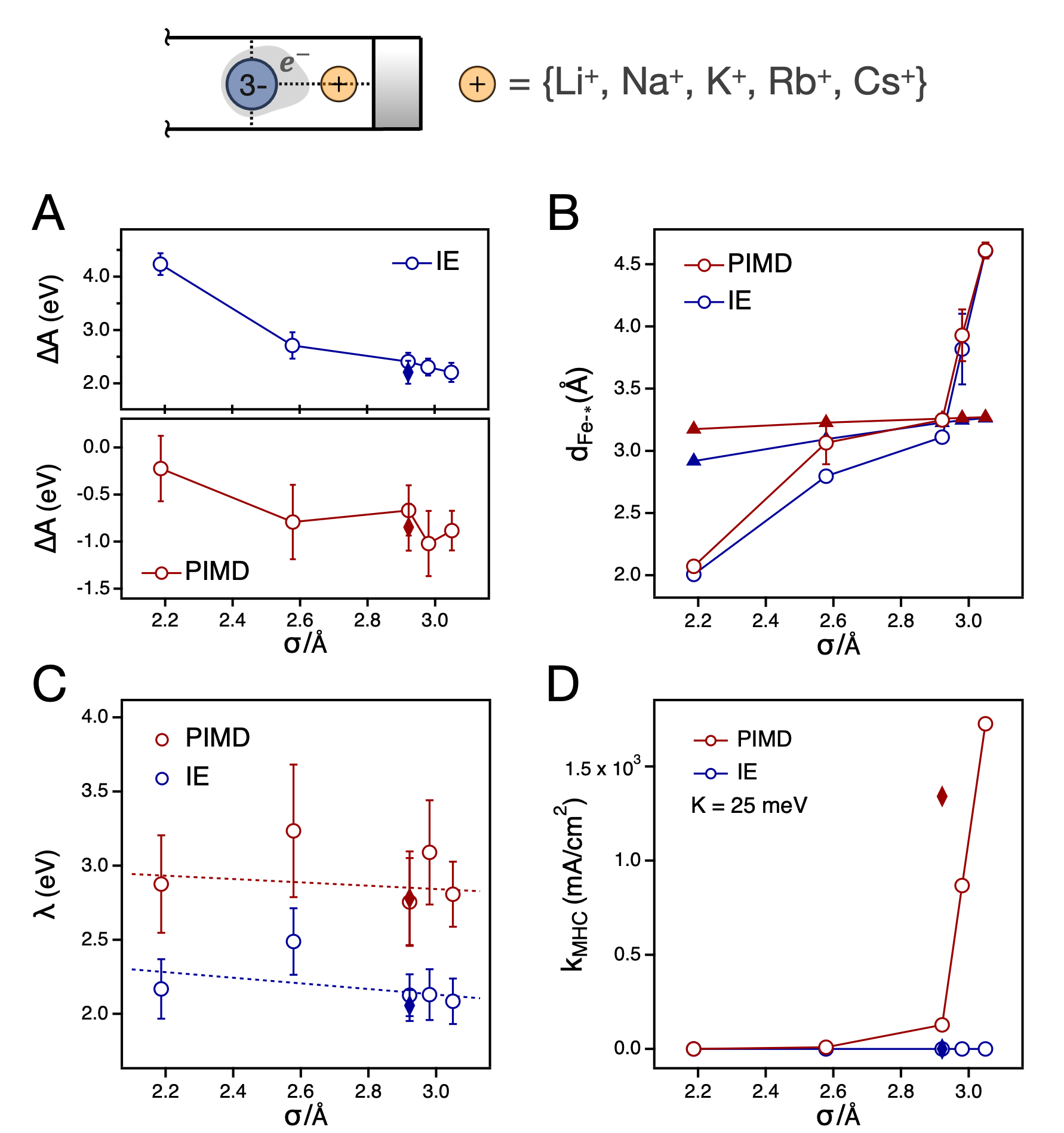}
\caption{Effect of bridging cation size on heterogeneous electron transfer properties. Results are shown as a function of cation diameter $\sigma$ where the bridging cation is positioned between the redox complex and the electrode surface (top schematic). Red symbols denote results from PIMD simulations whereas blue symbols indicate results from IE scheme. 
(A) Thermodynamic driving force $\Delta A$.
(B) Distance $d_{\text{Fe–*}}$ between the Fe atom and bridging cation (open circles) and N atom in redox complex (filled triangles). 
(C) Reorganization energy $\lambda$
(D) ET rate constant $k_{\mathrm{MHC}}$.
Filled diamond symbols in (A,C) indicate the results without bridging cation and cation species include Li$^+$, Na$^+$, K$^+$, Rb$^+$, and Cs$^+$.
}
\label{bcprop}
\end{figure}

To investigate the role of cation identity on interfacial electron transfer rate, we carried out simulations in which a cation with a varying ionic radius is constrained to reside between the redox species and the electrode, as illustrated in Figure~\ref{bcprop}.
In all cases, the redox species is constrained to reside at a distance from the electrode of $dz=8\mathring{\mathrm{A}}$.
No constraint was placed on the specific distance between the bridging cation and the redox complex, allowing it to equilibrate naturally within that region.
Ionic radii were selected to correspond with the series Li$^+$, Na$^+$, K$^+$, Rb$^+$, and Cs$^+$.
The average distance between the Fe atom at the center of the redox complex and the bridging cation (open circles) is shown in Fig.~\ref{bcprop}B, computed from both PIMD and IE simulations. 
Also shown are the average Fe–N distances within the anion complex (filled triangles), pointing to a transition in spatial organization: for small, high charge density cations such as Li$^+$, the bridging ion preferentially associates more closely with the redox center, even partially intercolating with the coordianting ligands, whereas for larger cations like Cs$^+$, reduced electrostatic attraction leads to a more peripheral positioning, indicating a weakening of direct interaction with the redox site. 

The thermodynamic driving force $\Delta A$, plotted in Fig.~\ref{bcprop}A, clearly reflects the transition behavior in bridging cation position. 
Among smaller cations, such as Li$^+$ and Na$^+$, $\Delta A$ exhibits a steep drop, which indicates a substantial shift in relative stability of the reduced state of the redox species.
As the cation size increases further, the changes in $\Delta A$ become more gradual, suggesting that the less strongly bound cations have a smaller influence on the relative stability of the reduced state of the redox species.
The influence of cation size is clearly manifested in the thermodynamic driving force $\Delta A$ across both PIMD and IE approaches, and similarly shapes the activation energy $\Delta A^{\ddagger}$ trends shown in Fig.S3. 

The reorganization energy shows relatively weak sensitivity to cation size (Fig.~\ref{bcprop}C).
This insensitivity is somewhat surprising given the pronounced differences in the positioning and stabilization effects described above.
One possible explanation for this insensitivity is that there is an interplay of counteracting effects.
For example, although Li$^+$ possesses higher charge density and would typically be expected to enhance solvent reorganization, its close proximity to the redox complex limits its influence on the surrounding solvent environment. On the other hand, larger cations like Cs$^+$ exhibit lower charge density but reside further from the redox center, allowing greater interaction with the surroundings. 
Such an interplay would align with the view that the primary role of the bridging cation is to modulate the stabilization of reactant and product states, rather than substantially altering the structural reorganization pathway. \cite{chen2016electric-0df, myint2021influence-8be}

Turning to the resultant ET rate constants $k_{\mathrm{MHC}}$, Fig.\ref{bcprop}D shows the computed ET rates from both the PIMD and IE schemes. Here, we observe that the ET rates obtained from the PIMD simulations increase markedly with cation size, primarily driven by the more favorable thermodynamic driving forces $\Delta A$ associated with larger cations. In contrast, the IE scheme yields negligibly small rates across all cations due to unrealistically high activation barriers, effectively masking any underlying size-dependent trends. 
The increasing trend in ET rates with cation size, as captured by the PIMD simulations, is in qualitative agreement with prior experimental observations, \cite{huang2021cationdependent-d65} providing additional support for the importance of considering the explicit details of the transferring electron in simulations of interfacial ET. 

Finally, to provide a reference for evaluating the bridging cation effect, we include results from a simulation without a bridging cation, marked by filled diamond symbols in each plot of Fig.\ref{bcprop}. These reference points correspond to simulations conducted with K$^+$ and a redox species–electrode separation of $dz = 8 \mathrm{\AA}$, as previously reported in Fig.\ref{V0prop}. Interestingly, in the absence of a bridging cation, we observe comparable reorganization energies but slightly more negative thermodynamic driving forces, resulting in higher ET rates. This outcome may stem from stronger direct electrostatic interactions between the redox complex and the electrode, which are otherwise partially screened by the presence of the cation. While these findings suggest that a cation-free interface can, in some cases, enhance ET kinetics, they more broadly highlight the sensitivity of ET behavior to the presence and size of intermediate ions in governing interfacial electron transfer processes.

\section{Conclusions}
\label{conclusion}

We have developed and applied a path integral molecular dynamics framework to model electron transfer at electrochemical interfaces, explicitly accounting for the quantum nature of the transferring electron. Using this framework, we construct Marcus parabolas from PIMD simluations to extract key ET properties—including reorganization energy, thermodynamic driving force, and activation energy—aligned with established physical understanding. By comparing this approach with a classical identity-exchange scheme, we demonstrated that the PIMD method provides robust estimates of these quantities, leading to ET rates that better reflect the underlying physics. 
Furthermore, our analysis of bridging cation effects reveals that the size and spatial positioning of intermediate ions can meaningfully influence ET thermodynamics, while the absence of a bridging cation results in a subtle change in thermodynamic driving force that contributes to a modest enhancement in the ET rate. 
These findings highlight the utility of the path-integral framework in capturing detailed interfacial behavior and emphasize the importance of explicitly modeling excess electrons when studying ET in complex electrochemical environments beyond classical approximations.


\vspace{5mm}


\bibliography{main}

\end{document}